\documentclass[sigconf]{acmart}
\usepackage[utf8]{inputenc}

\usepackage{amsfonts}
\usepackage{subfigure}

\usepackage{amsfonts}
\usepackage{siunitx}

\usepackage{tabularx}

\usepackage{color}

\usepackage{enumitem}

\usepackage{url}
\makeatletter
\def\url@leostyle{%
  \@ifundefined{selectfont}{\def\UrlFont{\sf}}{\def\UrlFont{\small\ttfamily}}}
\makeatother
\newcommand{\eat}[1]{}

\usepackage{xcolor}
\definecolor{light-gray}{gray}{0.9}

\usepackage{amsthm}

\newcolumntype{L}[1]{>{\raggedright\let\newline\\\arraybackslash\hspace{0pt}}m{#1}}

\title{Liquid Democracy in DPoS Blockchains}

\author{Chao Li}
\affiliation{%
  \institution{
  Beijing Key Laboratory of Security and Privacy in Intelligent Transportation \\
  Beijing Jiaotong University}
  \city{Beijing}
  \country{China}
}
\email{li.chao@bjtu.edu.cn}

\author{Runhua Xu}
\affiliation{%
  \institution{School of Computer Science and Engineering \\
  Beihang University}
  \city{Beijing}
  \country{China}
}
\email{runhua@buaa.edu.cn}

\author{Li Duan}
\affiliation{%
  \institution{
  Beijing Key Laboratory of Security and Privacy in Intelligent Transportation \\
  Beijing Jiaotong University}
  \city{Beijing}
  \country{China}
}
\email{duanli@bjtu.edu.cn}

\copyrightyear{2023}
\acmYear{2023}
\setcopyright{acmlicensed}\acmConference[BSCI '23]{The 5th ACM International Symposium on Blockchain and Secure Critical Infrastructure }{July 10, 2023}{Melbourne, VIC, Australia}
\acmBooktitle{The 5th ACM International Symposium on Blockchain and Secure Critical Infrastructure (BSCI '23), July 10, 2023, Melbourne, VIC, Australia}
\acmPrice{15.00}
\acmDOI{00.0000/0000000.0000000}
\acmISBN{000-0-0000-0000-0/00/00 }

\begin{document}
\fancyhead{}

\begin{abstract}

  Voting mechanisms play a crucial role in decentralized governance of blockchain systems.
  Liquid democracy, also known as delegative voting, allows voters to vote directly or delegate their voting power to others, thereby contributing to the resolution of problems such as low voter turnout.
  In recent years, liquid democracy has been widely adopted by Delegated-Proof-of-Stake (DPoS) blockchains and implemented successfully on platforms with millions of users.
  However, little is known regarding the characteristics and actual effectiveness of liquid democracy in decentralized governance.
  This paper explored for the first time the practical implementation of liquid democracy in DPoS blockchain systems.
  Using actual data collected from two major DPoS blockchains, EOS and Steem, our study compared and evaluated the participation of different types of users of DPoS blockchain systems in liquid democracy, as well as extracting and analyzing the delegation chains and networks formed during the process of liquid democracy within the systems.
  We believe that the findings of this paper will contribute to further studies on the design and implementation of liquid democracy and other voting mechanisms in decentralized governance.

\end{abstract}

\ccsdesc[500]{Security and privacy~Human and societal aspects of security and privacy}
\ccsdesc[500]{Networks~Network performance evaluation}

\keywords{Blockchain, Liquid Democracy, Decentralized Governance, Voting Mechanism, Delegated Proof of Stake, Web 3.0}

\maketitle

\section{Introduction}

Recent advancements in blockchain technology have rapidly revolutionized the perception of people of trust and decentralization~\cite{xu2021besifl,fan2019implement,lin2021measuring}.
Blockchain enables reliable and tamper-proof transactions without the necessity of intermediaries, laying a solid foundation for the growth of emerging fields such as Web 3.0 and Metaverse~\cite{duan2021metaverse,gai2022blockchain,gai2022blockchainUAC}.
One of the essential properties of blockchain systems is their decentralized governance, of which system updates, also known as \textit{forks}, are the most important component~\cite{ba2022fork,zhu2019blockchain}.
In traditional blockchains such as Bitcoin~\cite{nakamoto2008bitcoin}, proposals of \textit{forks} are typically negotiated off-chain by a small group of community members (e.g., large mining pools, core development teams) and enforced on-chain once an agreement has been achieved.
In this model, the selection of participants in decision-making lacks transparency, and the vast majority of users cannot participate in the decision-making process and can only passively accept or reject the results of decisions made by a few community members.
To make the decision-making process more transparent and decentralized, some recent Delegated-Proof-of-Stake (DPoS)~\cite{larimer2014delegated} blockchain systems, such as EOS~\cite{zheng2021xblock} and Steem~\cite{li2019incentivized}, are exploring the incorporation of voting mechanisms into the decision-making process. DPoS blockchain systems allow users to vote for a decision committee based on the weight of the coins they hold, thereby enabling any user to express their opinions and preferences.
However, due to many factors such as the lack of time or expertise, some users may choose not to vote, resulting in a low voter turnout ~\cite{kostadinova2003voter}.

Liquid democracy, also referred to as delegative voting, is a voting mechanism that offers voters two options: 
(1) vote directly for the committee;
(2) delegate their voting power to others.
The delegatees can make a selection between these two options once again.
If they vote directly, their votes would be weighted by the sum of their own voting power and delegated voting power.
Alternatively, they may choose to further delegate their voting power to others, creating a chain of delegation.
Thus, liquid democracy represents a more flexible voting mechanism that helps attract more users to vote by reducing the hidden cost of time and knowledge required to participate in voting.
The concept of liquid democracy has a rich history, dating back over 100 years. It was believed to be first envisioned by Charles Dodgson, the author of the novel Alice in Wonderland, in 1884~\cite{carroll1884principles}. 
Since then, liquid democracy has been explored in various settings, both online and offline. In recent years, it has been applied in many real cases, such as LiquidFeedback, an online voting platform operated by the German Pirate Party~\cite{kling2015voting}.
With the rise of blockchain technology, liquid democracy has gained considerable interest as a potential solution to the problem of low voter turnout in decentralized governance systems. 
It has been widely adopted by DPoS blockchains, where it has been successfully implemented on platforms with millions of users. 
Despite its widespread use, however, there is still much to learn about the characteristics and actual effectiveness of liquid democracy in decentralized governance. 

This paper explored for the first time the practical implementation of liquid democracy in DPoS blockchain systems.
Using data collected from two major DPoS blockchains, EOS and Steem, our study examined liquid democracy at three different levels.
We first extracted all the \textit{delegation operations} from the raw data, constructed the fine-grained relationship between each pair of delegator and delegatee, and investigated the participation of different types of users in liquid democracy on the two systems. 
We then aggregated the interrelated delegation operations into a \textit{delegation chain} and investigated the similarities and differences in characteristics of delegation chains across the two systems.
Finally, we constructed a \textit{delegation network} for EOS and Steem, which includes all delegation operations and delegation chains. We visualized a snapshot of the network and analyzed and compared its characteristics. 
Furthermore, we identified suspicious Sybil accounts within the delegation networks and compared their characteristics across the two blockchains.
Through cross-verification of results across three levels of analysis, we found that liquid democracy has been successfully adopted in both EOS and Steem, but its characteristics differ in the two systems. 
One possible reason for the differences is that EOS lacks the strong social attributes possessed by Steem, resulting in a flatter delegation network.
We believe that the findings of this paper will contribute to further studies on the design and implementation of liquid democracy and other voting mechanisms in decentralized governance.

\noindent \textbf{Contributions.}
In a nutshell, this paper makes the following key contributions:

\begin{itemize}[leftmargin=*]
\item We present the first in-depth analysis of the practical implementation of liquid democracy in DPoS blockchain systems. Our empirical analysis focuses on two major platforms, namely EOS and Steem.
\item We develop a model to examine liquid democracy at three different levels: delegation operations, delegation chains, and delegation networks. This model effectively captures the intricate relationships among the three levels and provides a comprehensive understanding of the voting mechanism in DPoS blockchains.
\item We identify the similarities and differences in characteristics of delegation chains across EOS and Steem. Our results offer insights into the varying features of liquid democracy between the two platforms.
\item We analyze the impact of social attributes on the structure of delegation networks. Our results explain the differences in liquid democracy implementation between EOS and Steem.
\item We identify suspicious Sybil accounts within the delegation networks and perform a thorough analysis of their characteristics across the two systems. Our analysis reveals patterns of behavior and potential vulnerabilities, which can facilitate the future development of countermeasures to enhance the robustness and security of liquid democracy implementations.
\item Overall, our findings contribute to the growing body of knowledge on the design and implementation of liquid democracy and other voting mechanisms in decentralized governance, which potentially inform future improvements in these systems.
\end{itemize}

\noindent \textbf{Organization.}
\textcolor{black}{
We start by introducing and modeling liquid democracy in Section~\ref{sec2}.
We then depict the two datasets utilized in this study in Section~\ref{sec3}.
In Section~\ref{sec4}, we investigate delegation operations in EOS and Steem by analyzing user participation, classifying delegators and delegatees, and monitoring their changes over time.
In Section~\ref{sec5}, we explore key properties of delegation chains in EOS and Steem, including intermediary growth and chain lengths.
In Section~\ref{sec6}, we focus on delegation networks by analyzing snapshots, identifying suspicious Sybil accounts, and comparing their characteristics across the two systems.
We discuss related work in Section~\ref{sec7} and conclude in Section~\ref{sec8}.}

\section{Liquid democracy}
\label{sec2}

In this section, we first deconstruct liquid democracy into three distinct levels, namely delegation operations, delegation chains, and delegation networks. 
We then present a model of liquid democracy for DPoS blockchain systems, which formally depicts the interconnected three levels. 


\subsection{Deconstructing liquid democracy}
The liquid democracy mechanism comprises three interrelated levels, namely delegation operations, delegation chains, and delegation networks. 
Together, they provide a comprehensive understanding of the liquid democracy mechanism in DPoS systems. 
Delegation operations serve as the fundamental building blocks, delegation chains consist of sequential operations, and delegation networks offer a holistic perspective on the delegation landscape.

\begin{itemize}[leftmargin=*]
  \item \textbf{Delegation operation.}
  A delegation operation represent the fundamental action in the liquid democracy mechanism. In liquid democracy, a user either votes directly for the committee using a voting operation or delegates their voting power to another user using a delegation operation. A delegation operation generates a pair of delegator and delegatee. The delegatee then decides whether to vote directly or delegate their voting power further.
  
  \item \textbf{Delegation chain.}
  A delegation chain emerges when a user delegates to another user, who in turn delegates their voting power, and so on. A delegation chain consists of multiple delegation operations, ending with a voter who votes directly. These chains include intermediaries, namely users who act as both delegators and delegatees within the chain.
  
  \item \textbf{Delegation network.}
  A delegation network is a comprehensive representation of all delegation operations and chains within a DPoS blockchain system. It can be visualized as a directed graph, where edges indicate delegation operations between users. Delegation networks can provide an overview of the entire voting landscape, which captures the flow of voting power and relationships among users in the system.
\end{itemize}




\subsection{Modeling liquid democracy}

\begin{figure}
  \centering
  {
      \includegraphics[width=1.0\columnwidth]{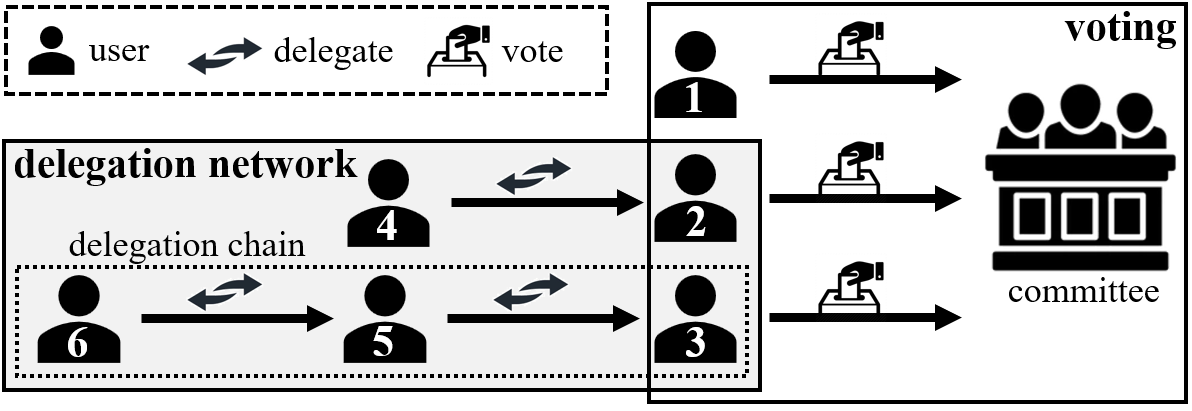}
  }
  \caption {\small A model of liquid democracy in DPoS blockchains.}
  \label{fig_pre} 
\end{figure}

After presenting the three levels of liquid democracy, we introduce a model of liquid democracy for DPoS blockchain systems. 
The model is shown in Figure~\ref{fig_pre}.
In this model, a finite set of users $U=\{u_1,u_2,...,u_n\}$ of a DPoS blockchain system want to vote for a committee $M$ to manage the blockchain system, including producing and validating blocks and handling proposals of \textit{forks}. 
The committee $M$ is a subset of a finite set of candidates $C=\{c_1,c_2,...,c_m\}$, namely users who apply to join the committee, and the size of $M$ is usually a fixed value (21 in EOS and Steem).
Next, we formalize the three levels.

\begin{itemize}[leftmargin=*]
  \item \textbf{Delegation operation.}
  In a DPoS blockchain system that employs liquid democracy, users can either vote directly for the committee or delegate their voting power to another user. 
  In the case that user $u_i$ delegates to another user $u_j$, $u_i$ and $u_j$ become the delegator and the delegatee, respectively. 
  This is defined as a \textit{delegation operation}, denoted by $d_i=u_j$.
  Delegators and delegatees are thus two subsets of users $U$, denoted by $D_s$ and $D_r$, respectively.
  If users vote directly, they are called \textit{voters} $V=\{v_1,v_2,...,v_l\}$.
  For instance, Figure~\ref{fig_pre} includes six users. Among them, $u_1$, $u_2$ and $u_3$ are voters while $u_4$, $u_5$ and $u_6$ have chosen to delegate to $u_2$, $u_3$ and $u_5$, respectively. 
  
  \item \textbf{Delegation chain.}
  If user $u_i$ delegates to user $u_j$, who in turn delegates to user $u_k$ and $u_k$ is a voter, this creates a \textit{delegation chain}, denoted by $d_i \to d_j$.
  A delegation chain includes multiple delegation operations and ends at a voter.
  It also includes at least one \textit{intermediary}, a user that is both a delegator and a delegatee.
  For instance, Figure~\ref{fig_pre} includes one delegation chain $d_6 \to d_5$, which has a length of two and involves a single intermediary $u_5$.
  
  \item \textbf{Delegation network.}
A \textit{delegation network} is formed by all the delegation operations and includes all the delegation chains.
A delegation network can be represented by a $G=(U, E)$, where $U$ is the set of users, and $E \subseteq U \times U$ is the set of directed edges. Each edge $(u_i, u_j) \in E$ indicates a delegation operation $d_i=u_j$, where user $u_i$ delegates voting power to user $u_j$.
In Figure~\ref{fig_pre}, $d_4=u_2$, $d_5=u_3$ and $d_6=u_5$ together forms a delegation network.
\end{itemize}

In Table~\ref{tab:notations}, we present a summary of the notations and their corresponding descriptions used throughout our discussion of the liquid democracy model for DPoS blockchain systems. 
This table serves as a reference for understanding the terminology and symbols employed in our analysis of delegation operations, delegation chains, and delegation networks.

\begin{table}[t]
  \centering
  \caption{Notations and Descriptions}
  \begin{tabularx}{\dimexpr\linewidth-0.3cm}{cX}
  \hline
  Notation & Description \\
  \hline
  $U$ & Finite set of users \\
  $u_1, u_2, ..., u_n$ & Individual users \\
  $M$ & Committee to manage the blockchain system \\
  $C$ & Finite set of candidates \\
  $c_1, c_2, ..., c_m$ & Individual candidates \\
  $d_i = u_j$ & Delegation operation (user $u_i$ delegates to user $u_j$) \\
  $D_s$ & Delegators (subset of users) \\
  $D_r$ & Delegatees (subset of users) \\
  $V$ & Set of voters \\
  $v_1, v_2, ..., v_l$ & Individual voters \\
  $d_i \to d_j$ & Delegation chain \\
  $G$ & Delegation network \\
  \hline
  \end{tabularx}
  \label{tab:notations}
\end{table}

\begin{figure*}
  \centering
  \subfigure[{\small Delegators and voters}]
  {
     \label{p1}
      \includegraphics[width=0.66\columnwidth]{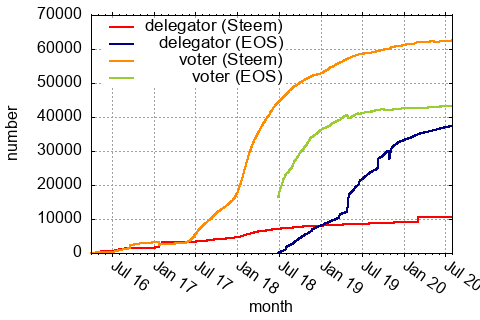}
  }
  \subfigure[{\small Delegatees and candidates}]
  {
    \label{p2}
      \includegraphics[width=0.66\columnwidth]{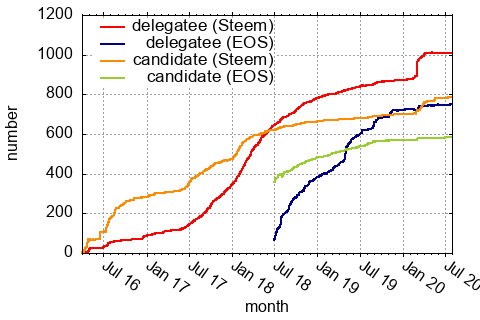}
  }
  \subfigure[{\small Delegation Representativeness Index}]
  {
    \label{p3}
      \includegraphics[width=0.66\columnwidth]{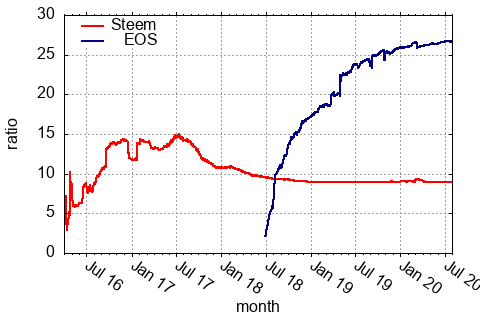}
  }
   \caption{Participation of different types of users in liquid democracy.}
  \label{participation}
\end{figure*}

\section{Datasets}
\label{sec3}

In this section, we introduce the two datasets utilized in this study.

\noindent \textbf{EOS.}
EOS is one of the most prominent second-generation blockchain projects that support smart contracts~\cite{wood2014ethereum}.
It was released in June 2018 and has frequently ranked in the top ten cryptocurrencies on coinmarketcap.com.
EOS has attracted a lot of attention from developers who are interested in using smart contracts to build decentralized applications.
These applications have rapidly gained a large number of users due to the high transaction throughput enabled by the DPoS consensus protocol.
We selected EOS as one of the study cases for liquid democracy due to its enormous number of users and transaction volume.
We obtained the EOSIO blockchain data produced between July 2018 and July 2020 from a recently published dataset~\cite{zheng2021xblock}.

\noindent \textbf{Steem.}
Steem is another highly successful DPoS blockchain system. 
It was released in March 2016 and was once ranked in the top thirty cryptocurrencies on coinmarketcap.com.
Steemit, the most well-known decentralized application based on Steem, is by far the most popular blockchain-based social media platform~\cite{li2019incentivized,guidi2021analysis}.
Steem has over a million users and billions of transactions, which makes it a suitable study case for liquid democracy.  
We collected the Steem blockchain data produced between April 2016 and July 2020 by utilizing their official API~\cite{li2021steemops}.

In the next three sections, we will utilize the datasets to investigate delegation operations, delegation chains and delegation networks, respectively.

\section{Delegation operation}
\label{sec4}

In this section, we first extract all the delegation operations and examine the participation of different types of users in liquid democracy in EOS and Steem.
We further classify delegators and delegatees based on attributes such as degree and voting power, and evaluate the changes of different categories of delegators and delegatees over time.


\subsection{Participation in liquid democracy}

In this part, we analyze the participation of four types of users in liquid democracy on EOS and Steem, including delegators, delegatees, voters, and candidates. We investigate the growth patterns of these user types over time and propose a Delegation Representativeness Index to measure the representativeness of delegatees. 

Figure~\ref{participation} depicts the participation of four types of users of EOS and Steem in liquid democracy, including delegators $D_s$ and delegatees $D_r$, as well as voters $V$ and candidates $C$.
In both EOS and Steem, delegators $D_s$ and delegatees $D_r$ are senders and recipients of the delegation operations.
Similarly, voters $V$ and candidates $C$ are senders and recipients of the vote operations.
Both types of operations converge voting power from tens of thousands of senders to hundreds of recipients, as illustrated in Figure~\ref{p1} and Figure~\ref{p2}, respectively.
The number of users in all four categories has grown steadily on both platforms over the four-year period from 2016 to 2020. 
However, the growth patterns vary across the categories and the platforms.

\begin{figure*}
  \centering
  \subfigure[{\small Delegatees based on degree}]
  {
     \label{eos_class_1}
      \includegraphics[width=0.66\columnwidth]{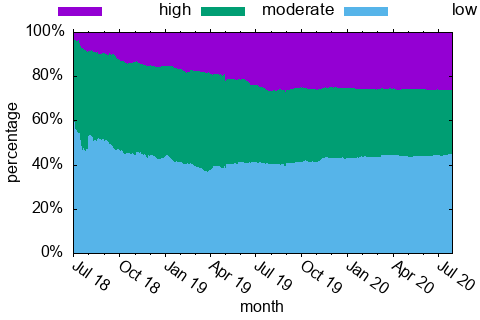}
  }
  \subfigure[{\small Delegatees based on voting power}]
  {
    \label{eos_class_2}
      \includegraphics[width=0.66\columnwidth]{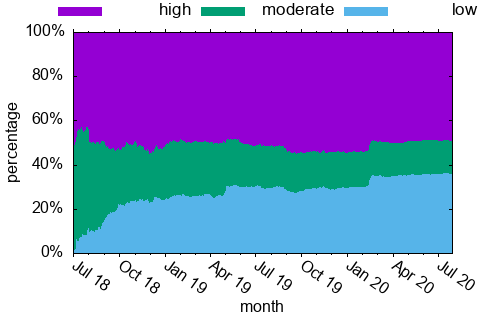}
  }
  \subfigure[{\small Delegators based on voting power}]
  {
    \label{eos_class_3}
      \includegraphics[width=0.66\columnwidth]{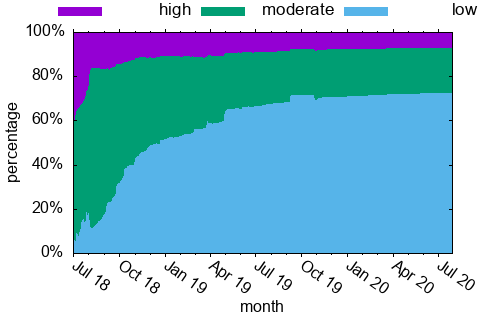}
  }
   \caption{Classification of delegatees and delegators in EOS.}
  \label{eos_class}
\end{figure*}

\begin{figure*}
  \centering
  \subfigure[{\small Delegatees based on degree}]
  {
     \label{steem_class_1}
      \includegraphics[width=0.66\columnwidth]{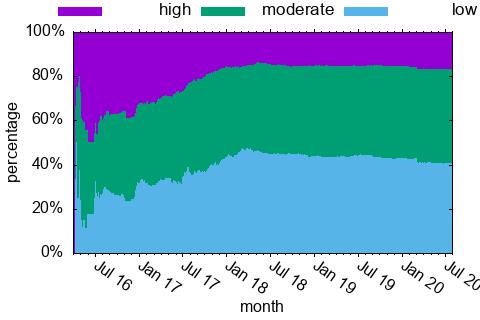}
  }
  \subfigure[{\small Delegatees based on voting power}]
  {
    \label{steem_class_2}
      \includegraphics[width=0.66\columnwidth]{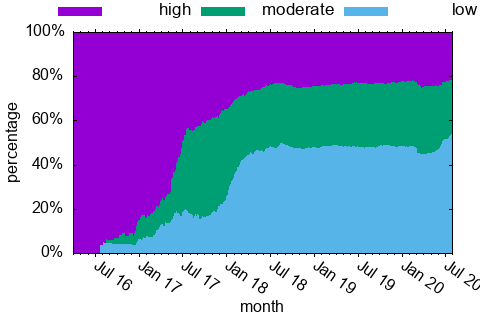}
  }
  \subfigure[{\small Delegators based on voting power}]
  {
    \label{steem_class_3}
      \includegraphics[width=0.66\columnwidth]{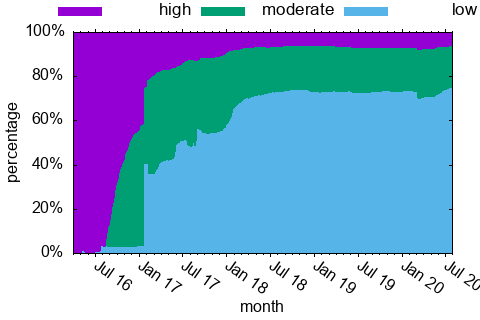}
  }
   \caption{Classification of delegatees and delegators in Steem.}
  \label{steem_class}
\end{figure*}

Concretely, Figure~\ref{p1} indicates that the number of delegators and voters in Steem was very close before Jul. 2017, but the surge in the number of voters after that made the gap between the two categories widen.
Conversely, in the case of EOS, within the first several months of its launch in 2018, the growth rate of voters exceeded that of delegators. Nonetheless, the growth of voters slowed down thereafter, while the growth of delegators remained steady. As a result, the gap between the two categories gradually decreased.
In spite of the fact that Steem was launched two years prior to EOS and both Steem and EOS initially supported liquid democracy, a notable difference in user behavior has emerged. 
While the majority of Steem users prefer to vote directly, the number of delegators in EOS has tended to surpass that of voters.
This phenomenon can be attributed to various factors, including differences in the ways of promoting liquid democracy and the ease of performing delegation operations. 
Additionally, the social nature of the user base of Steem, which tends to be more proactive and interactive, may also contribute to their preference for direct voting.

Then, as illustrated in Figure~\ref{p2}, the number of delegatees in Steem lagged behind that of candidates during the first two years but surpassed it in the subsequent two years. 
Similarly, the number of delegatees in EOS overtook that of candidates in the midway. 
These findings indicate that in both systems, the initial number of delegatees is lower than that of candidates, but the growth rate of delegatees is higher. 
This phenomenon is likely due to the lower requirements for becoming a delegatee than a candidate. Specifically, a candidate must actively apply for candidacy, while a user can become a delegatee by simply being designated as a recipient of a delegation operation by another user.

Based on the above discussion, we propose a \textit{Delegation Representativeness Index} $r$ to measure the representativeness of delegatees by integrating the numbers of delegators, delegatees, and voters. 
The index $r$ can be calculated as $r = a/b$, where $a=|D_s|/(|D_s|+|V|)$ assumes no liquid democracy and represents the proportion of delegators $|D_s|$ among all holders of voting power $(|D_s|+|V|)$, and $b=|V|/|D_r|$ represents the proportion of delegatees $|D_r|$ among all voters $|V|$ in liquid democracy. 
Intuitively, the index $r$ describes the fact that delegatees, who represent a proportion of $b\%$ with liquid democracy, actually represent the interests of delegators, who originally held a proportion of $a=rb\%$ without liquid democracy.
The higher the value of $r$, the higher representative the delegatees, and vice versa. 
Figure~\ref{p3} presents the index $r$ measured in Steem and EOS. As can be seen, the index $r$ in Steem increased from 5 to 15 in the first year but decreased to 9 in the second year and remained around 9 in the following two years. In contrast, the index $r$ in EOS continued to rise and reached about three times that of Steem in Jul. 2020. These findings suggest that the representativeness of delegatees in EOS is much higher than that in Steem.

In summary, our analysis revealed varying growth patterns across user categories and platforms, as well as a notable difference in user behavior emerging between EOS and Steem. 
While the majority of Steem users prefer to vote directly, the number of delegators in EOS has tended to surpass that of voters. 
We also introduced the Delegation Representativeness Index $r$ to measure the representativeness of delegatees and found that the representativeness of delegatees in EOS is significantly higher than that in Steem. 
These findings highlight the different dynamics and user preferences in liquid democracy across these two platforms.

\subsection{Classification of delegatees and delegators}

Next, we classify and analyze delegatees and delegators based on their degree and voting power on both EOS and Steem platforms. We divide delegatees into three categories according to their degree, and then further categorize delegatees and delegators into three distinct groups based on their voting power. 

We first divided delegatees into three categories according to their degree (i.e., delegated by how many delegators): (1) the low-degree delegatees with a single delegator; (2) the moderate-degree delegatees with delegators ranging from 2 to 10; (3) the high-degree delegatees with delegators more than 10.
The proportion of these three categories of delegatees among all delegatees is shown in Figure~\ref{eos_class_1} and Figure~\ref{steem_class_1} for EOS and Steem, respectively. 
It can be observed that in EOS, the proportion of high-degree delegatees increased from 3\% to approximately 26\% in the first year and remained stable in the second year, while the proportion of low-degree delegatees stabilized at around 40\% after experiencing a decline prior to April 2019.
In contrast, in Steem, the proportion of high-degree delegatees dropped from a peak of almost 50\% to below 20\% in less than two years, and has since remained around 16\%.
Overall, the proportion of the three categories of delegatees on both platforms tends to be gradually stable. Specifically, the proportion of low-degree delegatees on both EOS and Steem is around 40\%, indicating that a large number of delegatees are only delegated by a single delegator. Moreover, the proportion of high-degree delegatees in EOS is approximately 10\% higher than that in Steem, possibly due to the higher growth rate of delegators in EOS.

\begin{figure*}
  \centering
  \subfigure[{\small Intermediaries}]
  {
     \label{distance_1}
      \includegraphics[width=0.66\columnwidth]{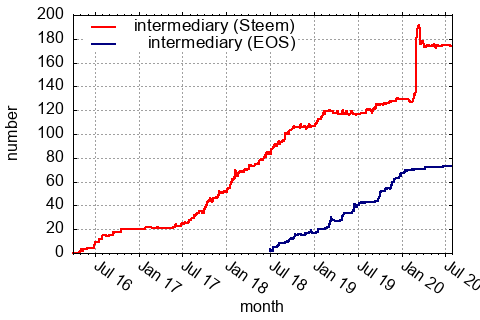}
  }
  \subfigure[{\small Lengths of delegation chains in EOS}]
  {
    \label{distance_2}
      \includegraphics[width=0.66\columnwidth]{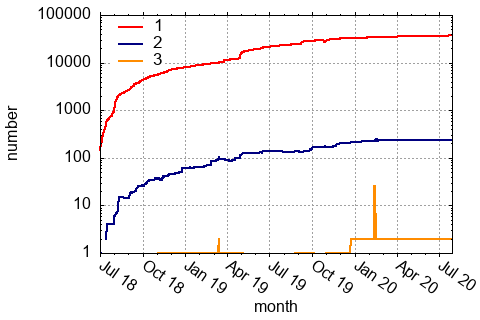}
  }
  \subfigure[{\small Lengths of delegation chains in Steem}]
  {
    \label{distance_3}
      \includegraphics[width=0.66\columnwidth]{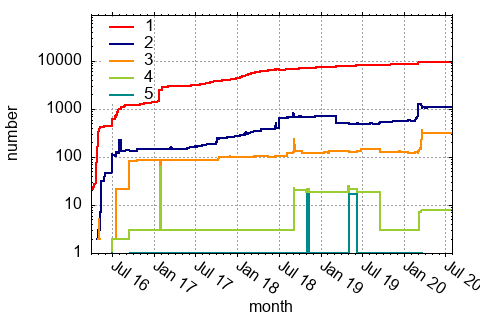}
  }
   \caption{Delegation chains.}
  \label{distance}
\end{figure*}

We then categorized delegatees into three distinct groups based on their voting power aggregated from delegators. The high-power delegatees were identified as those whose voting power ranks in the top 4\% of all voters, moderate-power delegatees as those with voting power ranking between the top 4\% and top 20\%, and low-power delegatees as those outside of the top 20\%. 
The results for EOS and Steem are presented in Figure~\ref{eos_class_2} and Figure~\ref{steem_class_2}, respectively. 
We also applied the same methodology to categorize delegators into three groups based on their own voting power, and the resulting groups are presented in Figure~\ref{eos_class_3} and Figure~\ref{steem_class_3} for EOS and Steem, respectively. 
As shown in these figures, approximately half of the delegatees in EOS are categorized as high-power, whereas less than 5\% of the delegators belong to the same group. 
This may highlight that liquid democracy results in the concentration of voting power among a small number of delegatees, which was previously dispersed among the delegators. The scattering of delegators' voting power and selected voting targets may lead to their votes having little impact on the election results. However, delegatees aggregate both voting power and voting targets, which enables low-power participants to compete with high-power participants in the election.
A similar concentration effect can be observed in Steem, but there are two main differences compared to EOS.
Firstly, the proportion of high-power delegatees in Steem was extremely high in the early stages but declined significantly in the following two years. This may be due to the fact that more new delegators selected different delegatees.
Secondly, although the proportions of the three categories of delegators in Steem were similar to those in EOS, the proportion of high-power delegatees was significantly lower in Steem. This may indicate that the aggregation of voting power in Steem is more dispersed than that in EOS.

In summary, our findings show that a large number of delegatees on both platforms are delegated by only a single delegator. Meanwhile, the proportion of high-degree delegatees in EOS is higher than that in Steem. 
We also observed that liquid democracy leads to the concentration of voting power among delegatees. This aggregation of voting power enables low-power participants to compete with high-power participants in the election. 
Moreover, the distribution of voting power among high-power delegatees and delegators in Steem is more dispersed than that in EOS, which suggests different patterns of voting power concentration between the two platforms.

\section{Delegation Chain}
\label{sec5}

In this section, we investigate delegation chains formed in EOS and Steem.
We first examine the growth of intermediaries, namely users that are both delegators and delegatees.
We then extract all delegation chains and explore their lengths.

Figure~\ref{distance_1} depicts the number of intermediaries in EOS and Steem over time.
It can be observed that the number of intermediaries, as a subset of delegatees, changes almost in proportion to the number of delegatees.
Specifically, the number of intermediaries accounts for about 10\% to 15\% of the total number of delegatees, slightly higher in Steem. 
Interestingly, in Mar. 2020, there was a sharp increase in the number of intermediaries and delegatees in Steem, which may be attributed to the influx of participants introduced to liquid democracy due to the hostile takeover that Steem was undergoing at that time~\cite{ba2022fork}.

Figure~\ref{distance_2} and Figure~\ref{distance_3} illustrate the changes in the lengths of all delegation chains in EOS and Steem over time. 
A delegation chain includes at least one intermediary, and its length is equal to the number of intermediaries plus one.
Therefore, the minimum length should be 2. 
In the figures, a length of 1 refers to an independent delegation operation, which means the delegator has not been delegated by other users.
It can be observed that the number of independent delegation operations in EOS is significantly higher than that in Steem. 
However, the number and lengths of delegation chains in EOS are much lower than those in Steem. 
Specifically, in EOS, except for a few chains with a length of 3, all the chains have a length of 2. 
In contrast, in Steem, there are almost a thousand chains with a length of 2, as well as over a hundred chains with a length of 3, and even a small number of chains with lengths of 4 and 5.
The underlying reason for these differences may lie in the more complex social network in Steem, which makes delegation operations among users more likely to be nested.
The lack of such a social network in EOS may result in flatter delegation relationships among users.

To sum up, our analysis in this section revealed that intermediaries constitute 10\%-15\% of delegatees, with Steem showing slightly higher proportions. We also found that EOS has more independent delegation operations, while Steem has longer and more numerous delegation chains. These differences can be attributed to Steem's more complex social network, resulting in nested delegation operations, while EOS exhibits flatter delegation relationships.



\begin{figure*}
  \centering
  \subfigure[{\small EOSIO}]
  {
     \label{graph_1}
      \includegraphics[width=1\columnwidth]{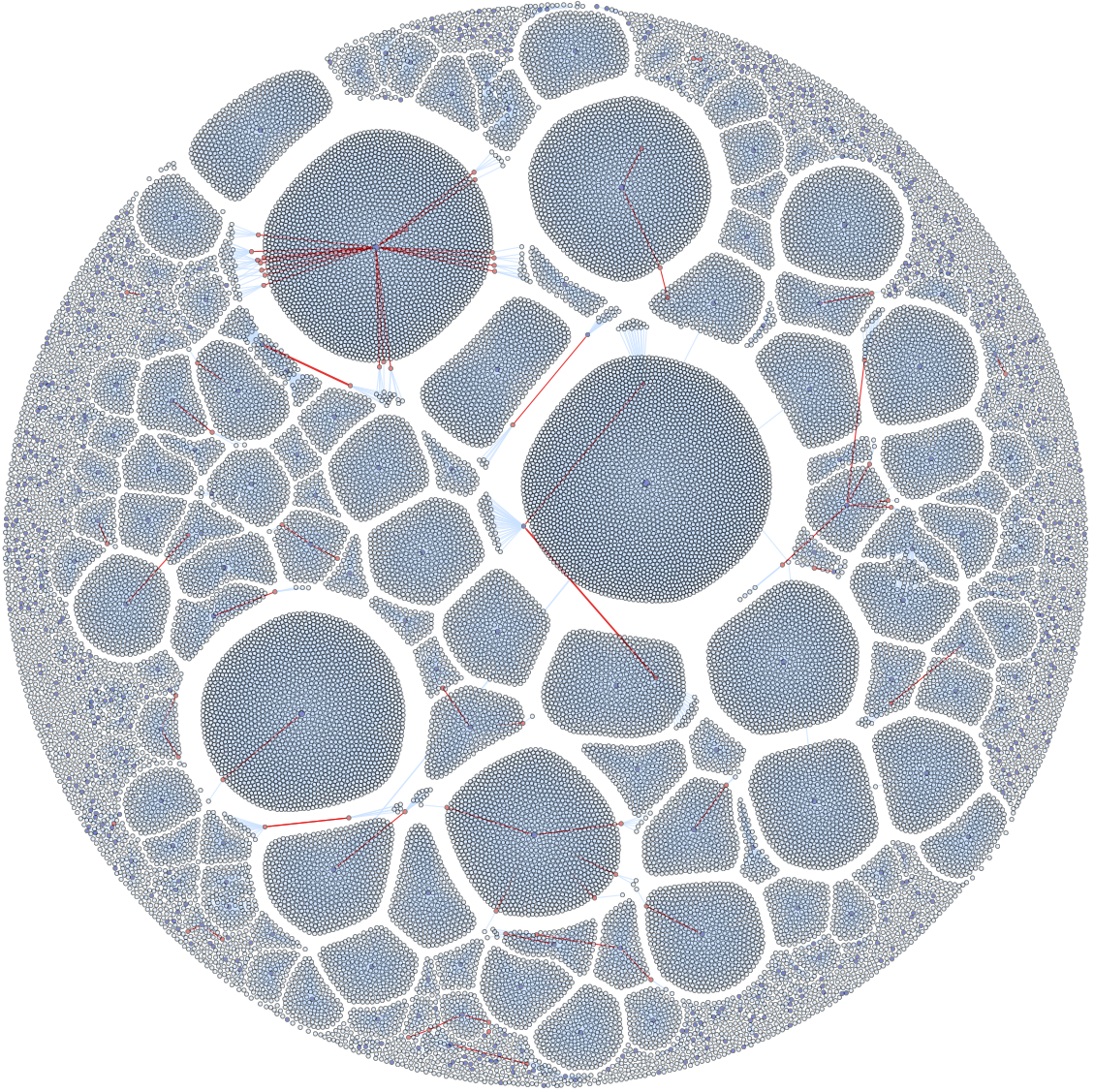}
  }
  \subfigure[{\small Steem}]
  {
    \label{graph_2}
      \includegraphics[width=1\columnwidth]{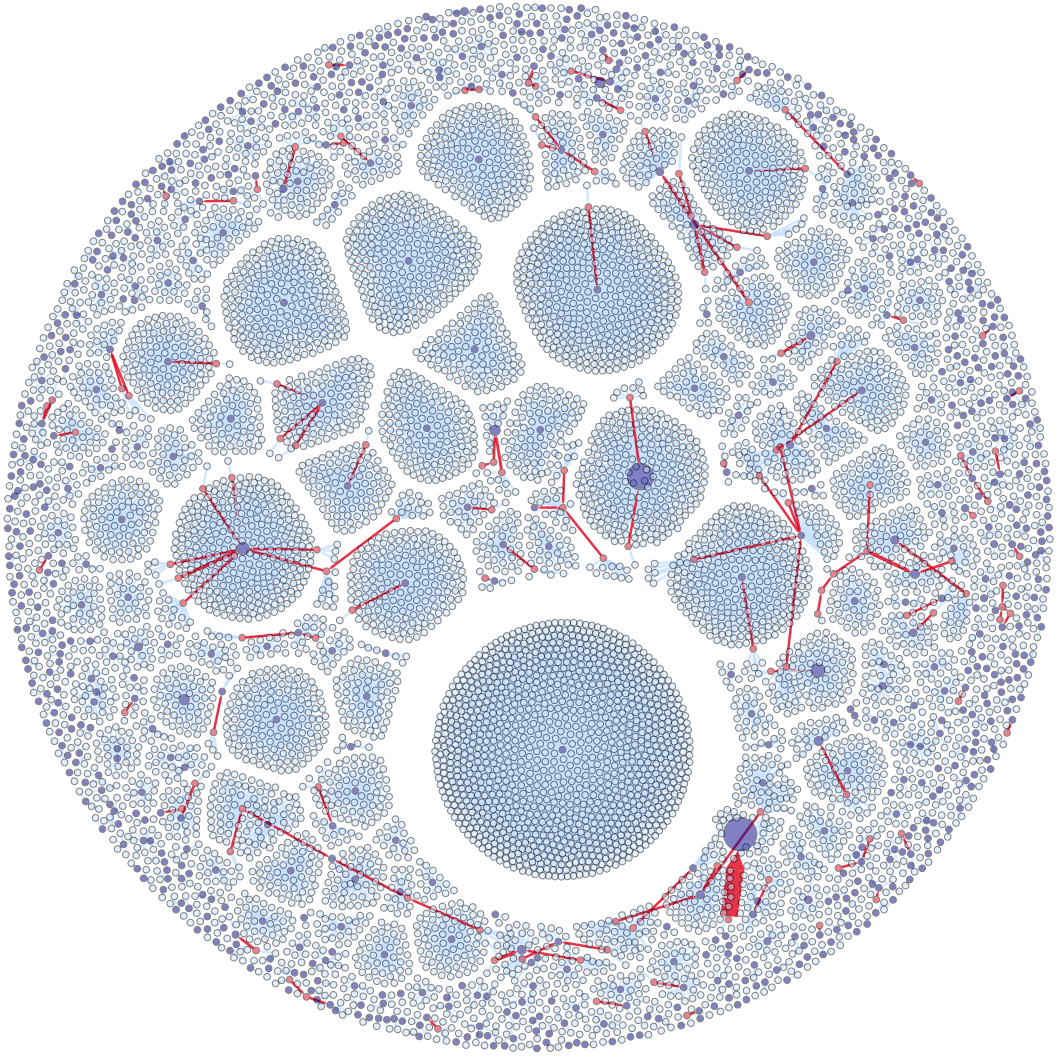}
  }
   \caption{Delegation networks captured on Feb. 14$^{th}$ 2020. The light blue and dark blue nodes represent delegators and delegatees, respectively, who are not intermediaries. The red nodes indicate intermediaries. The color of the edges corresponds to their source nodes, and the thickness of the edges corresponds to the voting power of their source nodes.}
  \label{graph}
\end{figure*}

\begin{table}
  \caption{Statistics of delegation networks.}  
  \begin{center}
  \begin{tabular}{|p{4.5cm}|p{1.2cm}|p{1.2cm}|}
  \hline
  {\textbf{Item}} & {\textbf{EOS}} & {\textbf{Steem}} 
  \\ \hline
      $\#$ of nodes & \textbf{35022} & 9983 \\
      $\#$ of edges (delegation operations) & \textbf{34263} & 9213 \\
      \hline \hline
      $\#$ of independent delegation operations & \textbf{34037} & 8402 \\
      $\#$ of delegation chains (length=2) & 222 & \textbf{558} \\
      $\#$ of delegation chains (length=3)  & 2 & \textbf{120} \\
      $\#$ of delegation chains (length=4)  & 0 & \textbf{3} \\
      $\#$ of delegation chains (length=5)  & 0 & \textbf{1} \\
      \hline \hline
      $\#$ of delegatees (degree=1) & 319 & \textbf{381} \\ 
      $\#$ of delegatees (1<degree<=10) & 224 & \textbf{374} \\
      $\#$ of delegatees (10<degree<=100) & \textbf{128} & 127 \\
      $\#$ of delegatees (degree>100) & \textbf{60} & 14 \\
      \hline
  \end{tabular}
  \end{center}
  \label{t1}
\end{table}

\section{Delegation Network}
\label{sec6}

In this section, we examine delegation networks formed in EOS and Steem and also identify sybil accounts suspected to be controlled by the same entity.
Our study utilizes two snapshots of the delegation networks captured in EOS and Steem on Feb. 14$^{th}$ 2020, a date close to but before the hostile takeover of Steem, so that more regular delegation operations could enrich the graphs.

\begin{figure*}
  \centering
  \subfigure[{\small EOSIO}]
  {
     \label{metric_1_E}
      \includegraphics[width=1\columnwidth]{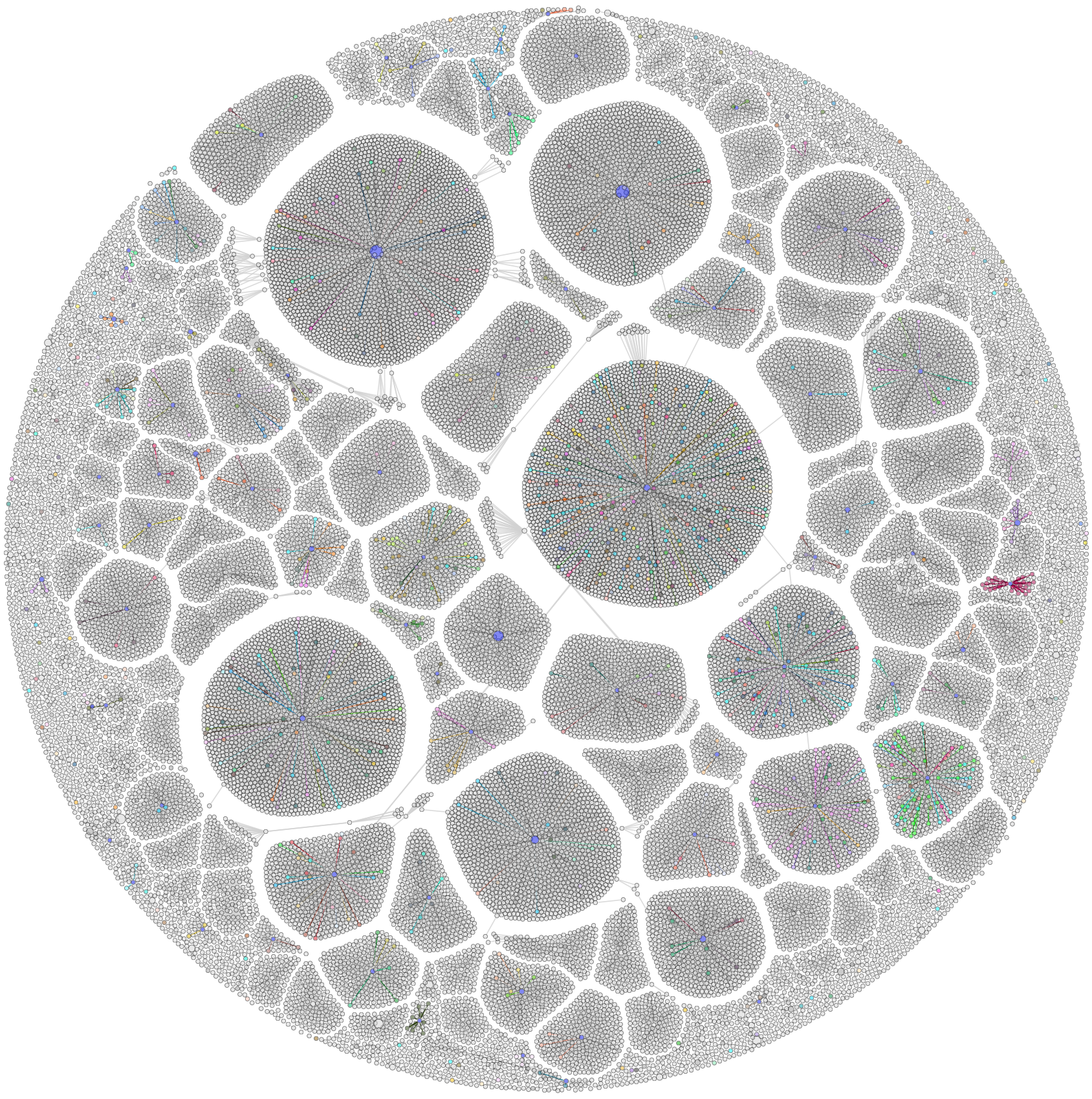}
  }
  \subfigure[{\small Steem}]
  {
    \label{metric_1_S}
      \includegraphics[width=1\columnwidth]{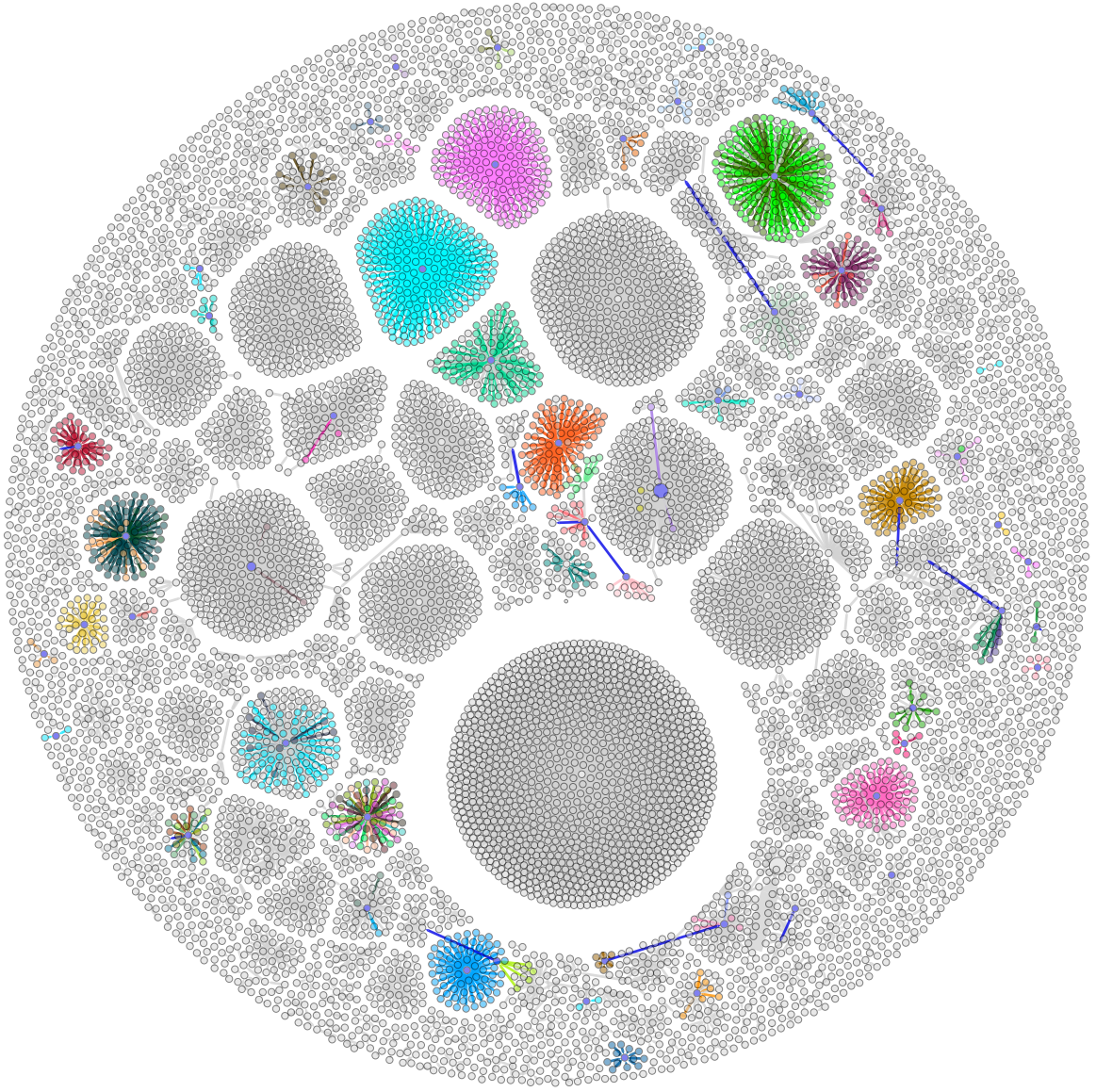}
  }
   \caption{Sybil accounts identified on Feb. 14$^{th}$ 2020. Nodes (delegators) with the same color within a single cluster are suspected to be sybil accounts controlled by a single entity.}
  \label{metric_1}
\end{figure*}

\subsection{Characterising Delegation Networks}

The two delegation networks are visualized in Figure~\ref{graph} and some statistics about the size of networks, length of delegation chains and degree of delegatees are presented in Table~\ref{t1}.
As can be seen, EOS has more nodes and edges, making its network much larger than that of the Steem network, which confirms the discussion in Section~4 about Figure~2. 
However, through an analysis of the lengths of delegation chains, it can be seen that over 99\% of delegation operations in EOS are independent.
There are only 224 delegation chains formed in EOS, of which only two reach a length of 3. 
In contrast, in Steem, approximately 84\% of delegation operations are independent.
There are 682 delegation chains, of which 124 reach a length of 3 and the longest reaches a length of 5. 
These phenomena confirm the discussion in Section~5 regarding Figure~5. 
Furthermore, an analysis of the degree of delegatees reveals that EOS has fewer low-degree and moderate-degree delegatees than Steem, while the number of high-degree delegatees, especially those with a degree higher than 100, far exceeds Steem. These findings confirm the discussion in Section~4 about Figures~3 and Figures~4.

Overall, the delegation network of EOS is larger, with more high-degree delegatees, but fewer delegation chains and shorter lengths of delegation chains. These findings are consistent with our previous analyses of delegation operations and chains. 
The delegation network in EOS is flatter, with the vast majority of delegators using independent delegation operations to directly delegate their voting power to voters. 
One possible reason for this phenomenon is that in EOS, only voters typically engage in personal promotion to potential delegators.
In contrast, due to its strong social networking attributes, all users in Steem can engage in personal promotion through social media platforms such as Steemit and can follow each other easily. 
Some users in Steem may delegate their voting power to their friends, who then delegate to their friends, making it easier to form delegation chains.

\subsection{Identifing Sybil Accounts}
Next, we identify delegators that are suspected to be Sybil accounts. 
We employ a straightforward yet effective heuristic for identification.
Specifically, we consider a group of delegators whose IDs consist of the same prefix (usually composed of letters or a combination of letters and numbers) and varying numerical suffixes to be Sybil accounts. 
For instance, in Steem, we discovered a group of 225 delegators with their IDs all starting with the same word "bitcoiner" followed by different numerical suffixes, resulting in IDs like "bitcoiner1", "bitcoiner2" and "bitcoiner3". 

The identification results are shown in Figure 7, where suspected Sybil accounts with the same prefix and different suffixes within the same cluster are represented using the same color. 
We found that, although the number of nodes in Steem is far less than in EOS, the number of suspected Sybil accounts in Steem, which is 1534, is considerably higher than in EOS, namely 795. 
Furthermore, as can be clearly seen in Figure 7, many clusters in Steem are entirely occupied by Sybil accounts controlled by one or two entities, while many clusters in EOS include Sybil accounts controlled by multiple entities, with a significantly higher proportion of non-Sybil accounts than Sybil accounts. 

Intuitively, this phenomenon may suggest that a small number of users in Steem were leveraging hundreds of Sybil accounts to concentrate voting power on a particular delegatee (probably themselves), which is not quite common in EOS. 
This intriguing behavior, although seemingly superfluous, can partially conceal the fact that a delegatee is merely supported by a single wealthy delegator, and create the appearance of broader support within the community. 
The underlying reason for this phenomenon may be that the Steem community resists oligarchic elections, which compels such users to partially hide their voting power. 
It is worth noting that the distribution of voting power among users in DPoS substantially affects the degree of decentralization in elections and governance~\cite{li2020comparison}. Consequently, this strategy of constructing Sybil accounts could also be used to mask the existence of substantial voting power controlled by a single user, which may eventually lead to a few entities dominating the decision committee and resulting in unfair governance.






\section{Related Work}
\label{sec7}

Liquid democracy has been widely explored in recent years. 
Kling et al. examined the emergence of super-voters in liquid democracy platforms. They investigated the distribution of power within Germany's Pirate Party and observed that super-voters use their power wisely, thus exhibiting a stabilizing effect on the system~\cite{kling2015voting}. 
Fan et al. proposed a fast algorithm to implement liquid democracy on the blockchain through Ethereum smart contract, which overcame the gas fee limitation and achieved an on-chain complexity of $O(log\ n)$ for processing each voting message, where $n$ is the number of voters~\cite{fan2019implement}. 
Zhang et al. presented a theory of power in delegable proxy voting systems, which includes an index to measure the influence of both voters and delegators~\cite{zhang2021power}. 
Fritsch et al. focused on the state of three prominent DAO governance systems on the Ethereum blockchain. They studied how voting power was distributed among governance token holders, delegates, proposals, and votes. They also investigated how voting power was used to influence governance decisions~\cite{fritsch2022analyzing}. 

DPoS-based blockchains have recently attracted significant attention. 
However, there is still a lack of research in analyzing the performance and security of DPoS blockchains. 
In 2019, Kwon et al. conducted a study on the degree of decentralization in various PoW, PoS, and DPoS blockchains~\cite{kwon2019impossibility}. 
Li et al. utilized the Shannon entropy in 2020 to quantitatively compare the degree of decentralization in Steem and Bitcoin~\cite{li2020comparison}. 
Huang et al. analyzed activities such as money transfers, account creation, and contract invocation in EOSIO and identified instances of bots and fraudulent activity~\cite{huang2020understanding}. 
In 2021, Guidi et al. conducted an analysis of the behaviors and social impact of committee members in Steem~\cite{guidi2021analysis}.
A study by Liu in 2022 characterized EOSIO and showed a gradual shift from decentralization to oligopoly~\cite{liu2022decentralization}. Tang et al. identified various types of user collusion behavior in Steem~\cite{tang2022identification}. 
To the best of our knowledge, this study represents the first empirical study to characterize liquid democracy in DPoS blockchains.

\section{Conclusion}
\label{sec8}

This paper provides an in-depth analysis of the practical implementation of liquid democracy in DPoS blockchain systems using actual data from two major blockchains. The findings suggest that liquid democracy has been successfully adopted in both EOS and Steem.
EOS has a more extensive delegation network and more high-degree delegatees, while Steem includes more and longer delegation chains. 
The differences may be attributed to the strong social networking attributes in Steem, which enable users in Steem to engage in personal promotion and easily form delegation chains. 
We believe these results contribute to further studies on the design and implementation of liquid democracy and other voting mechanisms in decentralized governance and could help uncover potential patterns of user behavior under different governance models.

\begin{acks}
This work was partially supported by the Beijing Natural Science Foundation under Grant No. M22039 and 4212008, the National Natural Science Foundation of China under Grant No. 62202038 and 62272031, and the Fundamental Research Funds for the Central Universities of China under Grant No. 2022JBMC007. 
Runhua Xu is the corresponding author (runhua@buaa.edu.cn).
\end{acks}

\renewcommand\refname{Reference}

\bibliographystyle{ACM-Reference-Format}
\urlstyle{same}

\bibliography{main.bib}

\end{document}